\documentclass[reprint,amsmath,amssymb, aps, pra]{revtex4-2}
\usepackage{amsmath,amssymb,physics,braket,subfloat,graphicx,subfigure,epstopdf,lastpage,hyperref}
\hypersetup{colorlinks=true, urlcolor=red, citecolor=blue, linkcolor=blue}
\usepackage{epstopdf}
\allowdisplaybreaks
\begin{document}
\title{Detecting nonclassicality and non-Gaussianity of a coherent superposed quantum state}
	\author{Deepak}
	\email{deepak20dalal19@gmail.com}
	\affiliation{Department of Mathematics, J. C. Bose University of Science and Technology,\\ YMCA, Faridabad 121006, India}
	\author{Arpita Chatterjee}
	\email{arpita.sps@gmail.com}
	\affiliation{Department of Mathematics, J. C. Bose University of Science and Technology,\\ YMCA, Faridabad 121006, India}
	\date{\today}
	\begin{abstract}

In this paper, we investigate the nonclassicality and non-Gaussianity of a coherent superposed quantum state (CSQS) which is obtained by applying a coherent superposition of field annihilation ($a$) and creation ($a^\dagger$) operators, $N(ta+ra^\dagger)$ to a classical coherent state $\ket\alpha$, where $t$ and $r$ are scalars with $t^2+r^2=1$. Such an operation, when applied on states having classical characters, introduces strong nonclassicality. We use different criteria to check the nonclassicality and non-Gaussianity of the considered quantum state. We first compute the Wigner function of CSQS. To study the nonclassicality of the considered state we further use (i) linear entropy (LE) (ii) Wigner logarithmic negativity (WLN) and (iii) skew information based measure. Relative entropy based measure is considered to analyze the variation in non-Gaussianity of CSQS. Finally, the dynamics of the Wigner function evolving under the photon loss channel is addressed to probe the effect of noise on nonclassicality as well as non-Gaussianity of CSQS.
\end{abstract}
	\maketitle
	\section{Introduction}
\label{intr}

In recent years, the nonclassicality and non-Gaussianity of a quantum state have received much importance as the investigation of fundamental structure of quantum states is crucial for current quantum technologies. The classification of a state as a classical or nonclassical one \cite{dp1} is an excellent example of such characteristics. This attention is well-deserved, as it has long been recognized that nonclassicality and quantum non-Gaussianity are two of the most important features of the quantum world, both of which can lead to quantum advantage \cite{ngncpsdfs1}. The quantification of nonclassicality provided by various quantum measures is also of great interest for experimental implementation of quantum optical technologies. Another state property, quantum non-Gaussianity \cite{dp2,dp3,dp4,dp5} is particularly important for various quantum technology based applications \cite{dp6,dp7,dp8}. Several distinct measures of nonclassicality and non-Gaussianity are applied to different quantum states to study these basic natures. P. Malpani et. al. \cite{png} performed a quantitative analysis of the nonclassical and non-Gaussian features for photon-added
displaced Fock state, as a test case, using a set of measures like entanglement potential, Wigner Yanese skew
information, Wigner logarithmic negativity and relative entropy of non-Gaussianity. K\"{u}hn et. al. \cite{dp} proved the usefulness of nonclassicality quasiprobabilities for certifying both quantum non-Gaussianity and the degree of nonclassicality. However, there is no unique approach for approximating the amount of nonclassicality or non-Gaussianity. In this work, we begin by defining the nonclassicality and non-Gaussianity, then discuss their significance in quantum technology and the urgency for quantifying these characteristics. We also check how the nonclassicality dominant parameters affect the non-Gaussianity.



Any quantum state $\rho$ can be represented in terms of the Glauber-Sudarshan $P$ function as \cite{ngncpsdfs2,ngncpsdfs3}
\begin{equation}
\rho = \int d(P(\alpha))\ket{\alpha}\bra{\alpha}
\end{equation}
The state is nonclassical if $P(\alpha)$ fails to be a probability distribution function. In other words, a nonclassical state cannot be represented in terms of a statistical mixture of coherent states. Nowadays the nonclassical states are most significant in the domain of quantum metrology, quantum computing, and quantum information processing \cite{ngncpsdfs4}. Major demand for enhancing device performance has raised, and it has been discovered that the desired enhancement can be achieved only by using quantum resources. Non-Gaussian states are those states which cannot be defined as a probabilistic combination of Gaussian states \cite{png20,dp}. Again, all the Gaussian states can be characterized by the initial two moments, which implies that the mean and covariance matrix of the Gaussian states provide their complete information. We can define a complex hull $\mathcal{G}$ with the classical probability distribution $P_{cl}(\lambda)$
\begin{equation}
\label{pcl}
\rho =\int{d(P_{cl}(\lambda))}\ket{\psi_\mathcal{G}(\lambda)}\bra{\psi_\mathcal{G}(\lambda)}
\end{equation}
in the Hilbert space $\mathcal{H}$. This set contains all the Gaussian states and a few non-Gaussian states. Interestingly, the non-Gaussian states obtained as the statistical mixtures of Gaussian states in the form \eqref{pcl} have limited applications due to their origin in classical noise \cite{dp,png22}. Thus quantum non-Gaussian states $\rho$ in $\mathcal{H}$ are the states which do not belong to the complex hull ($\rho\notin \mathcal{G}$). It is worth concentrating on the quantification of quantum non-Gaussianity in such a state as these states can be applied as more robust resources compared to the Gaussian states (\cite{png23,png24,png25}, and references therein). In particular, there are no-go theorems limiting the use of Gaussian operations and Gaussian states in entanglement distillation \cite{png26}, quantum error correction \cite{png27}, quantum computing \cite{png23}, and quantum bit commitment \cite{png28}. Use of non-Gaussian operations provides advantages in quantum computation \cite{png23}, quantum communication \cite{png29}, quantum metrology \cite{png24}, and so on. These promising applications motivated people to study about resource theory of non-Gaussianity \cite{png25}, and a number of measures of quantum non-Gaussianity (\cite{png30,png31,png32}, and references therein). It is worth to note that earlier quantum gravity and quantum information processing have been developed as two independent branches, but recently different significant and exciting results connecting these two fields have been reported (\cite{png33}, and references therein). It is observed that non-Gaussianity present in a state can be utilized to identify the signature of quantum gravity as only quantum theory of gravity can lead to non-Gaussianity \cite{png33}.

\begin{figure}[htb]
\centering
\includegraphics[scale=.55]{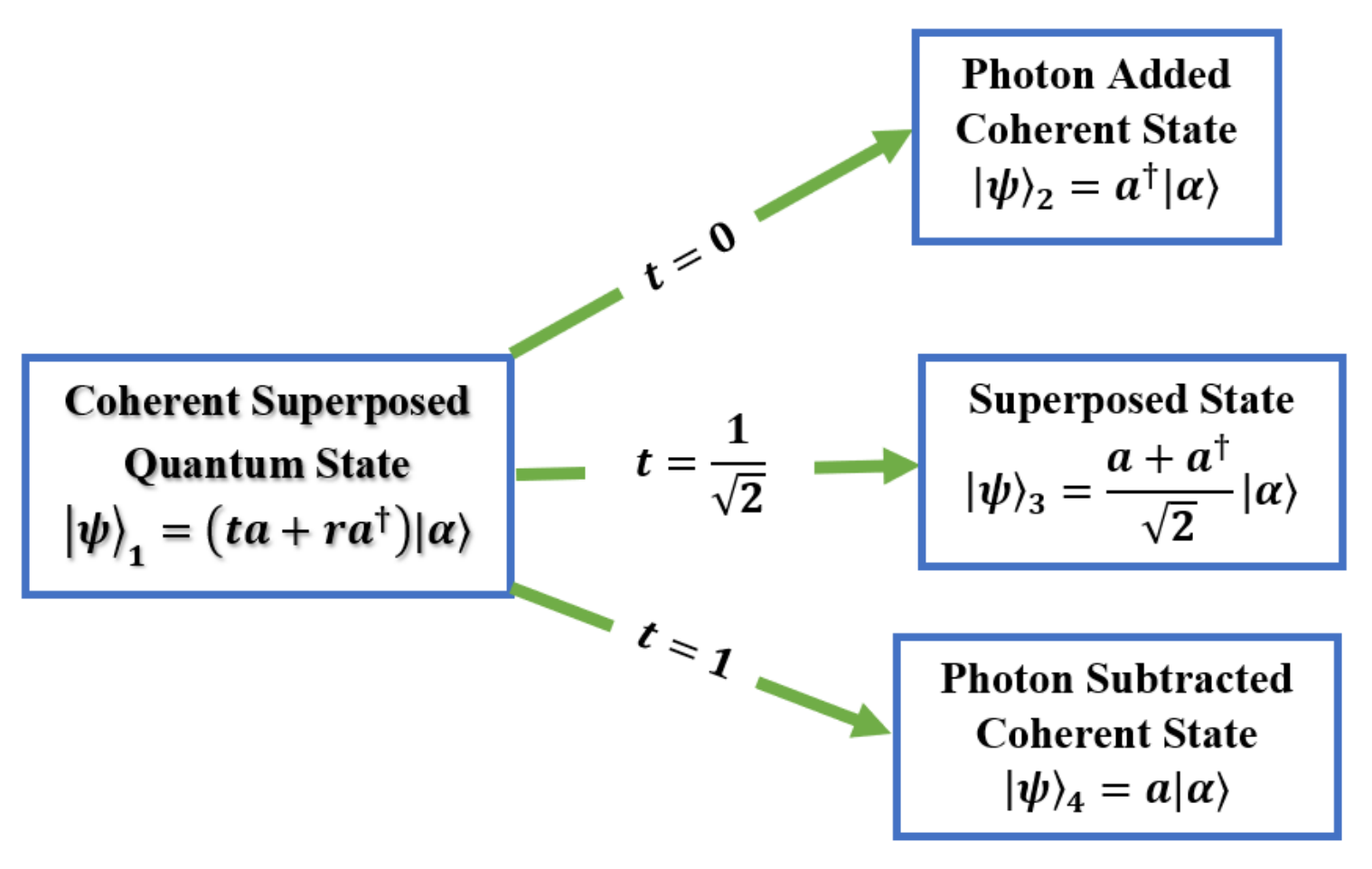}
\caption{Diagram for various limiting cases of CSQS i.e. for different values of $t$, CSQS reduces to different quantum states as shown in the figure.}
\label{figcsqs}
\end{figure}

In present study, we have considered a coherent superposed quantum states defined as a specific combination of bosonic creation and annihilation operators applied on the usual coherent state as $(ta+ra^\dagger)\ket{\alpha}$ with the standard relation $t^2+r^2=1$ \cite{rpslhn}. It is transformed to photon added (subtracted) coherent state for $t=0$ $(1)$ and superposed coherent state for $t=\frac{1}{\sqrt{2}}$ (see Fig.~\ref{figcsqs}).

This coherent superposition $(ta+ra^\dagger)$ acting on continuous variable system was first proposed by Lee and Nha for quantum state engineering \cite{rpslhn}. They have investigated how the superposed operation transforms a classical state to a nonclassical one together with emerging nonclassical
effects. They have also discussed the experimental feasibility of such a system by using beam splitters, parametric down converter and a nondegenerate parametric amplifier \cite{rpslhn1}. They have studied the case of generating an arbitrary superposition of number
states involving up to two photons, $C_0\ket{0}+C_1\ket{1}+C_2\ket{2}$, which can be used for quantum information processing. In a previous work, we have considered a generalized superposition of products of field annihilation ($a$) and creation ($a^\dagger$) operators of the type, $A = saa^\dagger+t{a^\dagger}a$ with $s^2+t^2=1$,
applied this SUP operator on coherent and thermal quantum states, and the states thus produced are referred as SUP-operated coherent state (SOCS) and SUP-operated thermal state (SOTS), respectively \cite{deepak1}. A comparative study between the higher-order nonclassical properties of SOCS and SOTS by using a set of nonclassicality witnesses (e.g., higher-order antiubunching, higher-order sub-Poissonian photon statistics, higher-order squeezing, Agarwal-Tara parameter, Klyshko's condition) was reported. Both of these work focused on the nonclassicality of the superposition states. In the present study, our aim is to investigate the non-Gaussian properties of the superposed state and to check if the nonclassicality enhancing parameters have any mere effect on the non-Gaussianity.

The structure of the paper is as follows: Sect.~\ref{qsuc} shows the detailed description of the CSQS state and its Wigner function. Sect.~\ref{mncg} describes different measures of nonclassicality and non-Gaussianity. The next section reports the Wigner function undergoing via a photon lossy channel. The article ends with a conclusion in Sect.~\ref{conclsn}.

\section{Detailed Description of CSQS and Its Wigner Function}
\label{qsuc}

\begin{figure*}[htb]
\includegraphics[scale=.9]{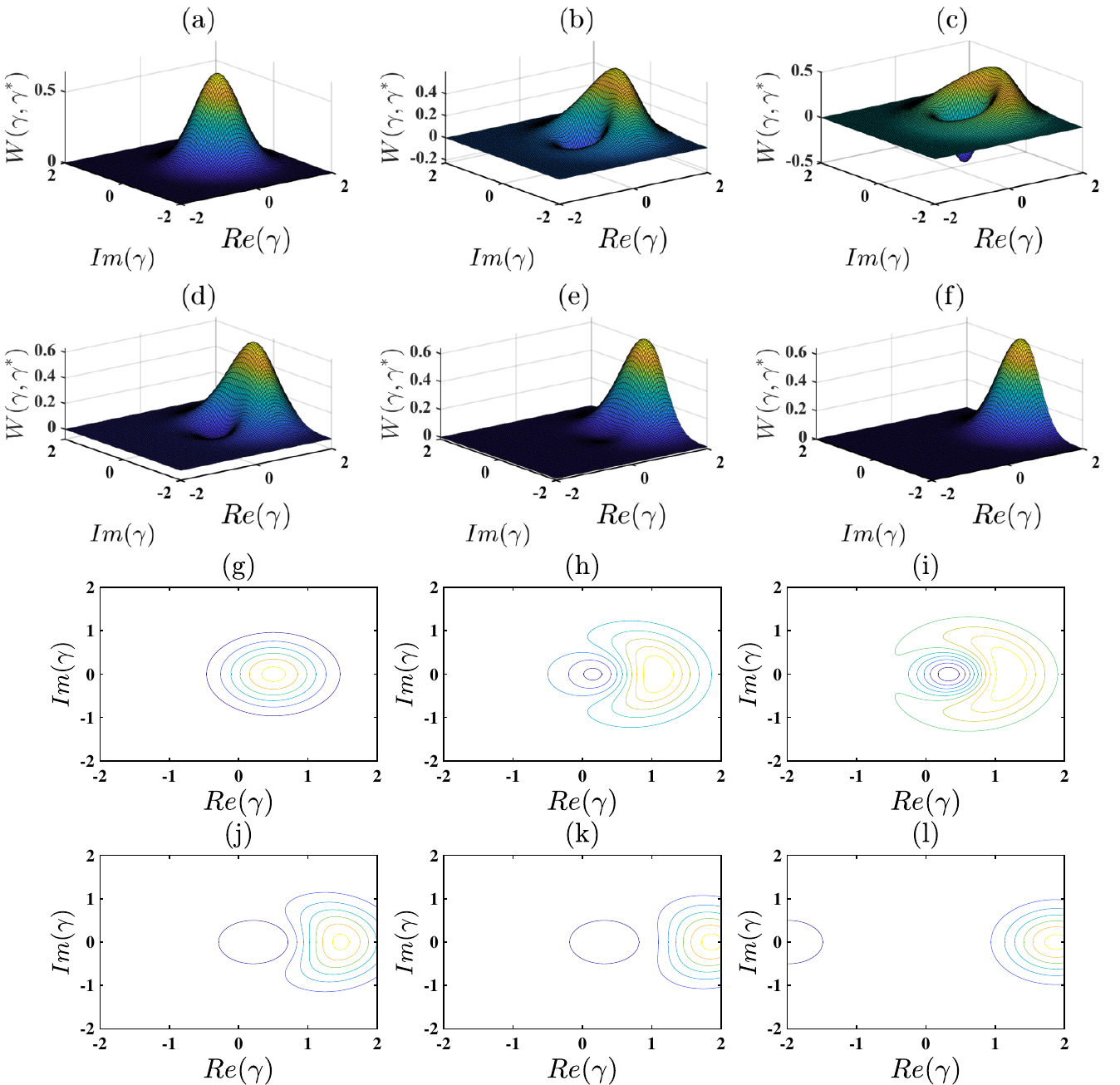}
\caption{A comparison of Wigner function with respect to real and imaginary $\gamma$ for $\alpha = 0.5$ and (a) $t=1$, (b) $t=1/\sqrt{2}$, (c) $t=0$ and for $t=0.5$ with (d) $\alpha = 1$, (e) $\alpha = 1.5$, (f) $\alpha = 1.75$, (g) to (l) are the corresponding contour plots. This figure shows non-Gaussian as well as non-classical behavior.}
\label{figwf}
\end{figure*}

Given a coherent state $\ket\alpha$ as an initial state, the output
state after operating the superposition operation is defined by $\ket\psi = N(ta+ra^\dagger)\ket\alpha$, where $N$ is the normalization constant as $N = \left\{|\alpha|^2+rt(\alpha^2+\alpha^{*2})+r^2\right\}^{-\frac{1}{2}}$. The Wigner function of the CSQS is derived as \cite{rpslhn}
\begin{equation}
\label{ewf}
W(\gamma,\gamma^*) = N^2[|t\alpha+r(2\gamma^*-\alpha^*)|^2-|r|^2]W_0(\gamma)
\end{equation}
where $W_0(\gamma) = \frac{2}{\pi}\exp(-2|\gamma-\alpha|^2)$ is the Wigner function of the initial coherent state $\ket{\alpha}$. This Wigner function is plotted with respect to real and imaginary $\gamma$ for different values of the state parameters $\alpha$ and $t$ in Fig.~\ref{figwf}.

The existence of nonclassical and non-Gaussian features in the coherent superposed quantum state can be studied in terms of the Wigner function. The negative region of the Wigner function indicates the presence of nonclassicality. In addition to that, it was proven by Hudson \cite{hud} that any pure quantum state having a non-negative Wigner function is necessarily Gaussian. It may be noted from \eqref{ewf} that the Wigner function has a Gaussian form for $t=1$ as it corresponds to the Wigner function of the photon-subtracted coherent state. It is clear from Fig.~\ref{figwf} that the non-zero values of the displacement parameter can influence the position of the non-Gaussian peak of the state for all values of $t$. Moreover, switching $t$ from 1 to 0 leads to the non-Gaussianity of the Wigner function having a Gaussian factor. The negative region of the Wigner function in Fig.~\ref{figwf} exhibits the nonclassical as well as the non-Gaussian features of the CSQS. It is also observed that with a decrease in values of $t$ keeping $\alpha$ fixed at 0.5, the negativity of the Wigner function increases.


\section{Different Measures of nonclassicality and non-Gaussianity}
\label{mncg}

In this section, we show how the non-Gaussianity and nonclassicality present in CSQS vary with respect to different state parameters (i.e. displacement ($\alpha$), superposition ($t$) parameters). This study is very much relevant in the context that such an exercise may help to identify the suitable state parameters for an
engineered quantum state which is to be used to perform a quantum computation, communication or metrology task that requires
nonclassical and/or non-Gaussian states. For the quantification of nonclassicality, the closed form analytical expressions of an
entanglement potential (to be referred as linear entropy potential), skew information based measure, and Wigner logarithmic
negativity are obtained in this section. Further, Wigner logarithmic negativity and relative entropy of non-Gaussianity are computed as measures of non-Gaussianity.

\subsection{Linear Entropy Potential}

Asboth \cite{pmng46} proposed a new measure of nonclassicality based on the fact that if a single-mode nonclassical (classical) state is inserted from one input port of a beam-splitter (BS) and a vacuum state is embedded from the other end, the output two-mode state must be entangled (separable). As a result, the nonclassicality of the input single-mode state (other than the vacuum state inserted in the BS) can be indirectly measured by using an entanglement measure. There are a variety of quantitative entanglement measures available which can be used to estimate the nonclassicality of the input single-mode state. When a measure of entanglement following Asboth's approach is used to assess the single-mode nonclassicality, the corresponding entanglement measure is referred to as the entanglement potential in correspondence with Asboth's terminology. In particular, if concurrence (linear entropy) is used to quantify the nonclassicality of the single-mode input state by measuring the entanglement of the two-mode output state departing from the BS, the nonclassicality criteria is called concurrence potential (linear entropy potential) \cite{pmng17}.

\begin{figure}[htb]
\centering
\includegraphics[scale=.27]{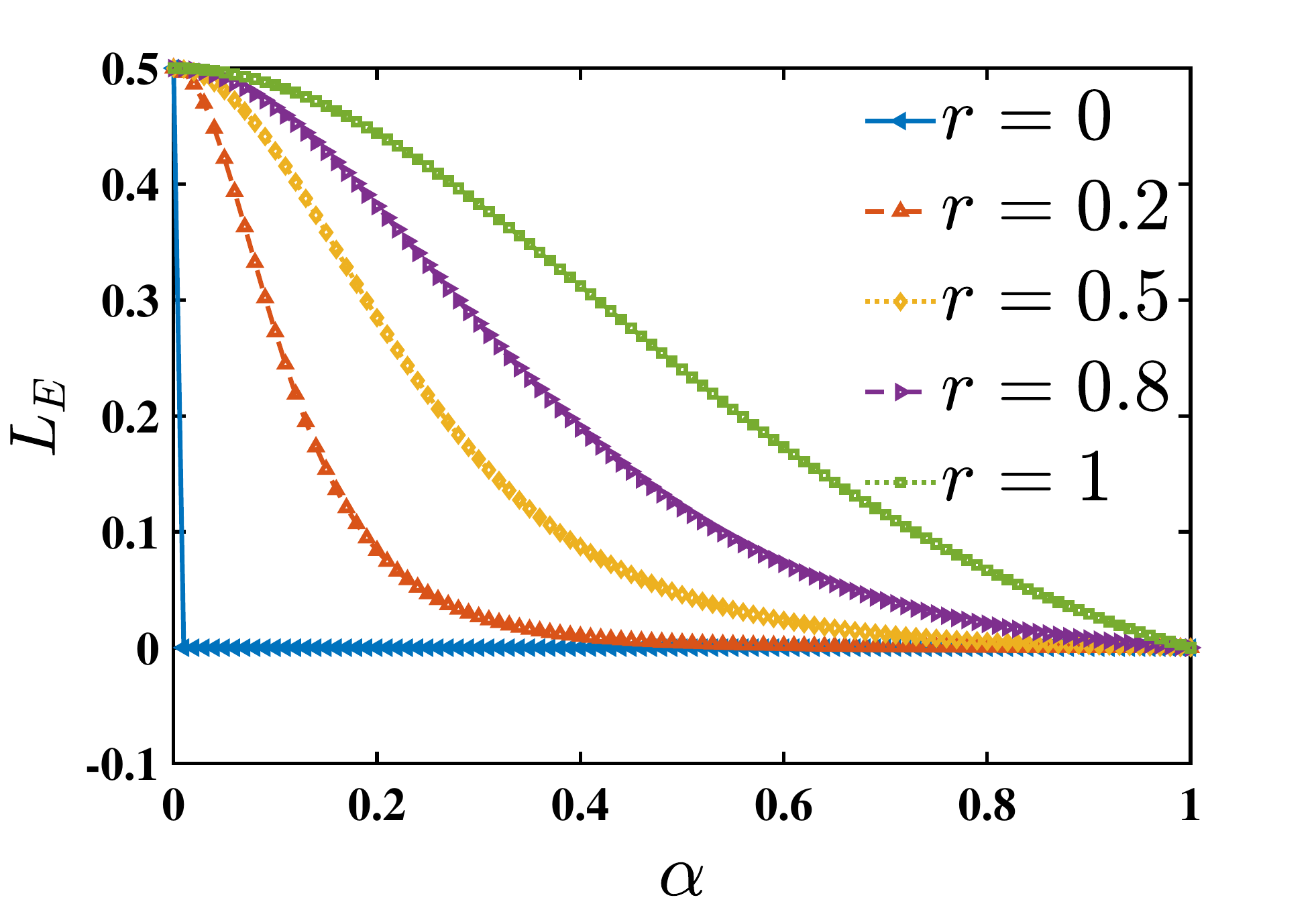}
\caption{Variation of LE with respect to the state parameter $\alpha$ for different values of $r$. It shows that nonclassicality is increasing with $r$ and decreasing with $\alpha.$}
\label{figep}
\end{figure}

The linear entropy \cite{png} for a bipartite state $\rho_{AB}$ is defined in terms of the reduced subsystem as
\begin{equation}
\label{ile}
L_E = 1-\text{Tr}(\rho_B^2)
\end{equation}
where $\rho_B$ is the partial trace of $\rho_{AB}$ over the subsystem $A$. Therefore, for a maximally entangled (separable) state, $L_E = 1$ $(0)$ and obtains a non-zero value for an entangled state in general \cite{png47,png48}. To compute the linear entropy potential, we would require the post-BS state $\rho_{AB}$ that originated if CSQS and vacuum states are mixed at a beam-splitter. The expression for the two-mode output state with Fock state at one port and vacuum at the other is as follows:
\begin{equation}
\label{beam}
\ket{n}\otimes\ket0=\ket{n,0}\,\xrightarrow{BS}\,\frac{1}{2^{n/2}}\sum_{j=0}^n \sqrt{{n\choose j}}\ket{j,n-j}
\end{equation}
This equation can be used to find the post-BS density matrix $\rho_{AB}=\ket{\phi}_{AB} {}_{AB}{\bra{\phi}}$ as follows: the coherent superposed quantum state in terms of number state basis is expressed as
$$\ket{\psi}=\sum_{p=0}^\infty \left(c'_{1p}\ket{p}+c'_{2p}\ket{p+1}\right)$$
with $c'_{1p}=N\,te^{-\frac{|\alpha|^2}{2}}\frac{\alpha^{p+1}}{\sqrt{p!}}$ and $c'_{2p}=N\,re^{-\frac{|\alpha|^2}{2}}\sqrt{p+1}\frac{\alpha^{p}}{\sqrt{p!}}$. Then using \eqref{beam}
\begin{align}
\label{phiep}
{\ket{\phi}}_{AB}\nonumber &= \ket{\psi}\otimes\ket{0}\\ &\,\xrightarrow{BS}\, \sum_{p=0}^\infty \left(c_{1p}\sum_{j=0}^p\sqrt{p\choose j}\ket{j,p-j} \right. \nonumber\\ &+ \left. c_{2p}\sum_{j=0}^{p+1}\sqrt{p+1\choose j}\ket{j,p+1-j}\right)
\end{align}
where $c_{1p}=\frac{c'_{1p}}{2^{p/2}}$ and $c_{2p}=\frac{c'_{2p}}{2^{(p+1)/2}}$. The simplified form of LE can be obtained as (see Appendix~\ref{aep} for detailed calculation)
\begin{align}
\label{SEP}
L_E & = 1-N^4\left[t^4|\alpha|^4 + t^3r\left\{2\alpha^2|\alpha|^2 + 2\alpha^{*2}|\alpha|^2 \right\} \right. \nonumber\\& +\left. t^2r^2 \left\{2|\alpha|^2(1+2|\alpha|^2) + \alpha^4 + \alpha^{*4} \right\} \right. \nonumber \\ & + \left.
tr^3\left\{2\alpha^2(1+|\alpha|^2) + 2\alpha^{*2}(1+|\alpha|^2)\right\} \right. \nonumber\\& +\left. r^4(1+5|\alpha|^2+2|\alpha|^4)/2 \right]
\end{align}
This LE is plotted in Fig.~\ref{figep} with respect to the state parameter $\alpha$ and for different values of $r$.
The variation in the amount of nonclassicality with respect to the state parameters is shown in Fig.~\ref{figep}. The nonclassicality is detected through the linear entropy potential. It is clear that LE increases as $r$ increases from 0 to 1 and decreases with increasing $\alpha$. The maximum value of LE obtained for the coherent superposed quantum state is 0.5 approximately.

\subsection{Criteria Based on Skew Information}

S. Luo et. al. \cite{png49} introduced the skew information based measure of nonclassicality in the context of Wigner Yanase skew information \cite{png50} in 2019. For a pure state $\rho$, this measure is defined as
\begin{equation}
\label{isk}
N(\rho) = \frac{1}{2} + \langle a^\dagger a\rangle - \langle a^\dagger\rangle\langle a\rangle
\end{equation}
$N(\rho)$ denotes the quantum coherence of $\rho$ with respect to the annihilation and creation operators, and is relatively easy to calculate. This measure is based on averages, which takes numerical value $\frac{1}{2}$ for classical coherent state and $n+\frac{1}{2}$ for mostly nonclassical $n$ photon Fock
state. Thus any state $\rho$ with $N(\rho)>\frac{1}{2}$ is nonclassical. However, this is only a one-sided condition for nonclassicality as it fails for some Gaussian mixed states \cite{png49}.

\begin{figure}[htb]
\centering
\includegraphics[scale=.27]{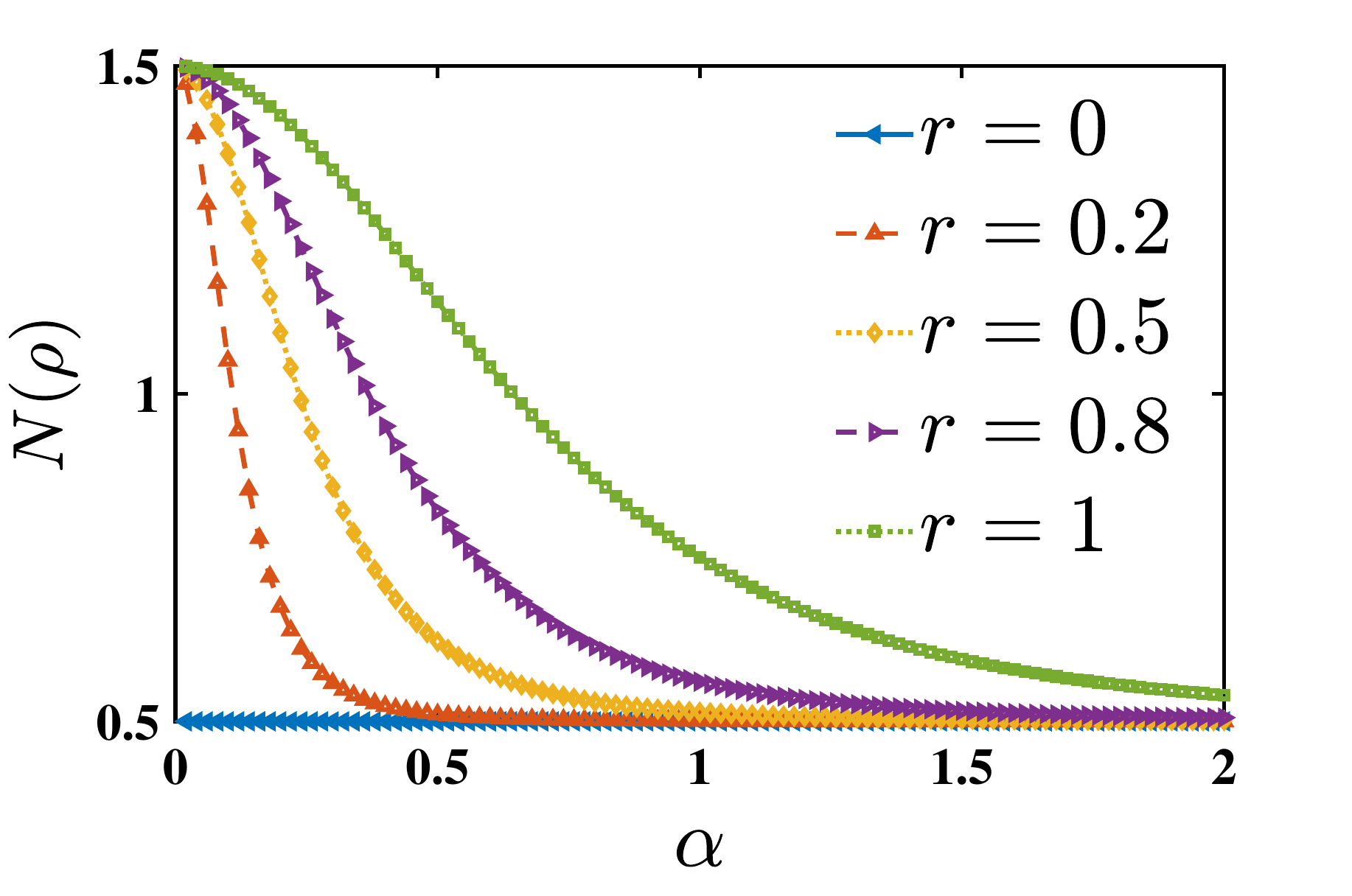}
\caption{Variation of $N(\rho)$ with respect to the state parameter $\alpha$ and for distinct values of $r$. It also shows behavior similar to linear entropy.}
\label{figsi}
\end{figure}

In the case of CSQS, $N(\rho)$ can be calculated using the general expectation of $a^{\dagger p}a^q$ as follows:
\begin{align}
\label{eadman}
\langle{a^{\dagger m}a^n}\rangle & = N^{2}\alpha^{* m-1}\alpha^{n-1}\Big[|\alpha|^4 + rt\Big\{(m+|\alpha|^2)\alpha^2 \\&+(n+|\alpha|^2)\alpha^{* 2}\Big\} +r^2\Big\{mn+(m+n+1)|\alpha|^2\Big\}\Big], \nonumber
\end{align}
The skew information based measure for CSQS can be simplified as
\begin{align}
N(\rho)& =\frac{1}{2} +N^{2}\Big[|\alpha|^4 + rt(1+|\alpha|^2)\Big\{\alpha^2+\alpha^{* 2}\Big\} \nonumber \\&+r^2\Big\{1+3|\alpha|^2\Big\}\Big] -N^{4}|\alpha|^{-2}\Big[|\alpha|^4 +  rt\Big\{(|\alpha|^2)\alpha^2 \Big. \nonumber\\ & \Big.+(1+|\alpha|^2)\alpha^{* 2}\Big\} +r^2\Big\{2|\alpha|^2\Big\}\Big]\Big[|\alpha|^4 + rt\Big\{(1+|\alpha|^2)\alpha^2 \Big. \nonumber\\ & \Big.+(|\alpha|^2)\alpha^{* 2}\Big\}+ r^2\Big\{2|\alpha|^2\Big\}\Big]
\end{align}
and is plotted in Fig.~\ref{figsi}.
It can be observed that the skew information based criteria exhibits nonclassicality of CSQS as the value of $N(\rho)~\geq~1/2$ for distinct parametric values. Further $N(\rho)$ is increasing (decreasing) with an increase in $r$ ($\alpha$). That means skew information based measure behaves similar to linear entropy. Also, the value of $N(\rho)$ lies between 0.5 and 1.5 approximately. Figure~\ref{figsi} shows that the nonclassicality expressed by the skew information based measure cannot be enhanced by increasing the coherent state parameter. It can also be seen that with an increase in values of $r$, nonclassicality increases.

\subsection{Logarithmic Negativity of Wigner Function}

We have seen in Fig.~\ref{figwf} that the Wigner distribution diagnoses the nonclassical as well as the non-Gaussian nature of the CSQS. This motivated us to quantify the amount of nonclassicality and non-Gaussianity using the volume of the negative part of the Wigner function. A measure of nonclassicality, named Wigner logarithmic negativity, estimates the amount of non-Gaussianity present in the CSQS as the negative values of the Wigner function also witness the non-Gaussianity of CSQS. The Wigner logarithmic negativity is defined as \cite{png25}
\begin{equation}
\label{iwln}
W = \log_2\left(\int{d^2\gamma}|W(\gamma,\gamma^*)| \right)
\end{equation}
where the integration is performed over the complete region $\mathcal{R}^{2n}$ of phase-space.

The integration in \eqref{iwln} is computed to study the effect of different parameters on
the Wigner logarithmic negativity of CSQS as (see Appendix \ref{awln} for details)
\begin{equation}
\label{aewln}
W = \log_2[N^2((t+r)^2\alpha^2-5r^2)]
\end{equation}

\begin{figure}[htb]
\centering
\includegraphics[scale=.27]{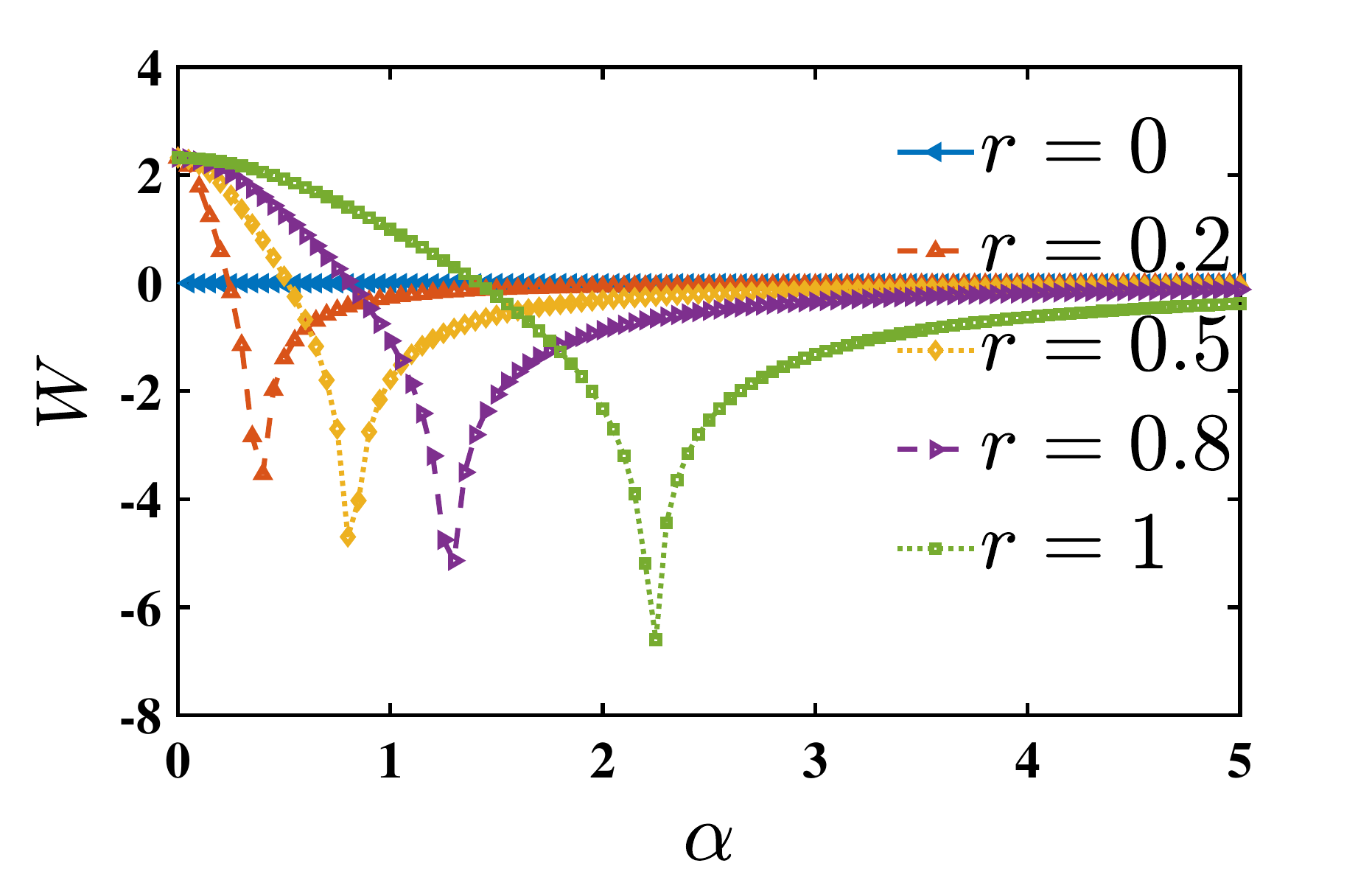}
\caption{Variation of $W$ with respect to state parameter $\alpha$ for distinct values of $r$. It also shows behavior similar to previous criteria and its negative value shows that the volume of non-classicality and non-Gaussianity shown by the Wigner function is less than one.}
\label{figwln}
\end{figure}

Fig.~\ref{figwln} shows that $W$ has a negative value in some region of $\alpha$. Further, it can be noticed that $W$ first decreases up to a fixed value of $\alpha$ and then increases. That means the amount of both nonclassicality and non-Gaussianity decreases up to a fixed value of $\alpha$ and then increases for all values of $r$ and $\alpha$. 

Therefore, all these criteria show that CSQS is nonclassical as well as non-Gaussian, and the amount of nonclassicality increases as $\alpha$ increases in a specific region. In the next section, we find another criteria to determine the amount of non-Gaussianity.

\subsection{Relative Entropy for Measuring the Amount of non-Gaussianity}

\begin{figure}[htb]
\centering
\includegraphics[scale=.27]{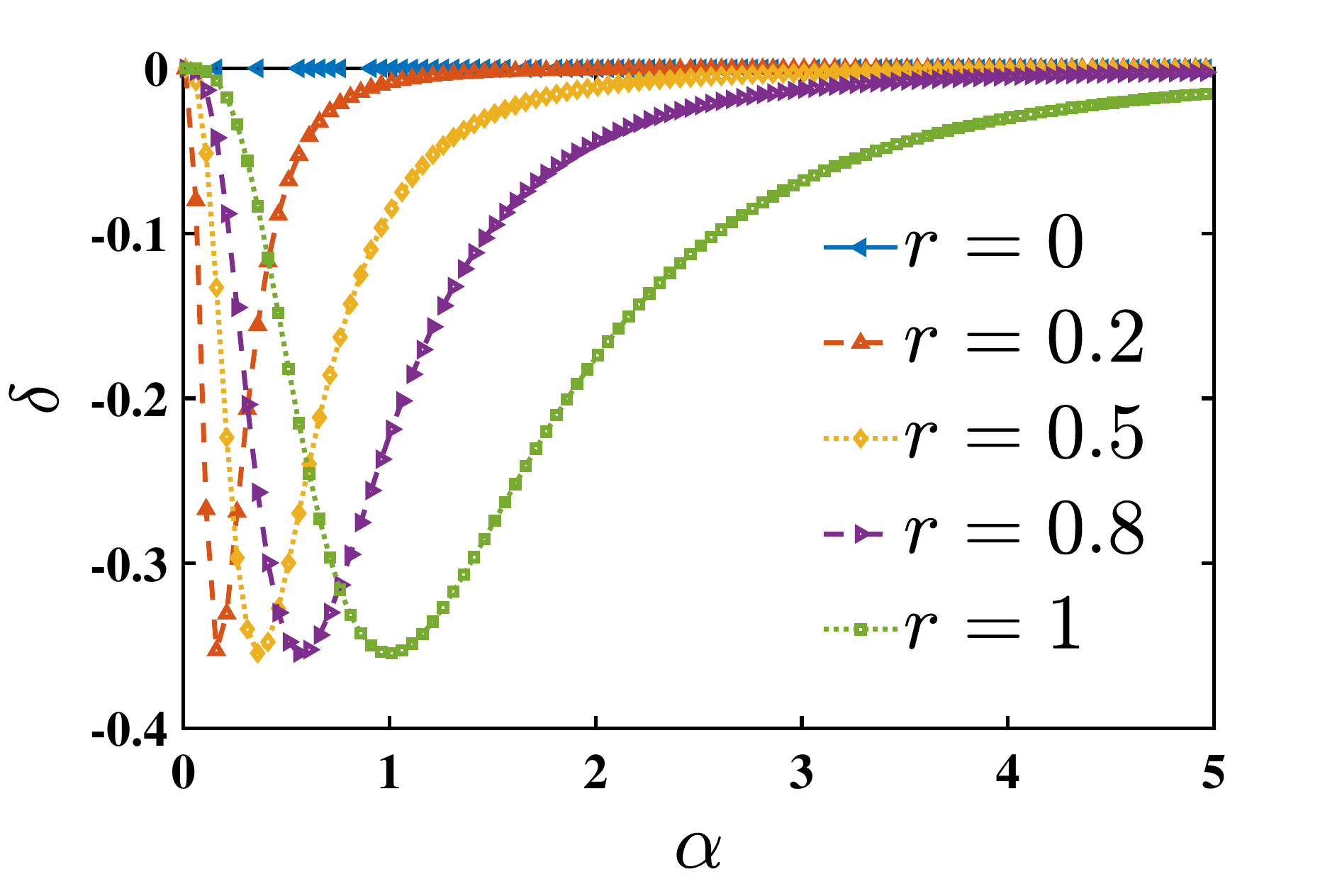}
\caption{Variation of $\delta$ with respect to $\alpha$ for different values of $r$. The behavior of $\delta$ is similar to $W$.}
\label{figmng}
\end{figure}

In the previous section, the non-Gaussianity is quantified by using Wigner logarithmic negativity and it is found that in case of CSQS, variations in
the amount of non-Gaussianity with the state parameters of interest is analogous to that of nonclassicality as quantified through
different measures. This is justified as the set of Wigner negative states is a subset of non-Gaussian states \cite{png25}. In what follows,
we reinvestigate this feature by using relative entropy of non-Gaussianity. The relative entropy of non-Gaussianity is defined in terms of its covariance matrix consisting of the first and second-order moments as \cite{png}
\begin{equation}
\label{matrix}
\sigma = \left[
\begin{matrix}
\sigma_{pp} & \sigma_{qp} \\  \sigma_{qp} &\sigma_{pp}
\end{matrix}
\right]
\end{equation}
where $\sigma_{pq}=\langle{pq+qp}\rangle -2\langle{p}\rangle\langle q\rangle$ and $p=\frac{a+a^\dagger}{\sqrt{2}},\,q=\frac{a-a^\dagger}{i\sqrt{2}}$. Using \eqref{eadman}, all the elements of the covariance matrix $\sigma$ can be calculated, and thus the relative entropy measure of non-Gaussianity reduces
to \citep{png}
\begin{equation}
\label{rmng}
\delta[\ket\psi]=S(\tau_G) = h(\det(\sqrt{\sigma}))
\end{equation}
with $h(x)=\frac{x+1}{2}\log_2\left(\frac{x+1}{2}\right)-\frac{x-1}{2}\log_2\left(\frac{x-1}{2}\right)$. The relative entropy of non-Gaussianity is plotted in Fig.~\ref{figmng}.

Fig.~\ref{figmng} shows similar nature of the relative entropy of non-Gaussianity for coherent superposed quantum state as shown by Wigner logarithmic negativity. There is a region of $\alpha$ where $\delta$ is decreasing first then it is increasing. Hence, we can conclude that amount of both nonclassicality and non-Gaussianity first decreases and then increases for all values of parameters. Further, all these parameters show that nonclassicality and non-Gaussianity increase with the increase in $r$ for keeping $\alpha$ fixed.

%

\section{Wigner Function of CSQS evolving under  Photon Loss Channel}
\label{wlc}

The interaction between a quantum system and its surroundings motivates quantum to classical transition. So, the observed nonclassical and non-Gaussian features are expected to decay due to the evolution of CSQS under the lossy channel. In particular, the temporal evolution of a quantum state $\rho$ over the lossy channel can be studied using the LGKS master equation \cite{png57} given by
\begin{equation}
\label{lgksme}
\frac{\partial\rho}{\partial t} = \kappa(2a\rho a^\dagger -a^\dagger a\rho - \rho a^\dagger a)
\end{equation}
where $\kappa$ is the rate of decay. Analogously, the time evolution of the Wigner function at time $t$ in terms of the initial Wigner function of the state evolving under a lossy channel \cite{png58} can be defined as
\begin{equation}
\label{dwlc}
W(\zeta,t) = \frac{2}{T}\int\frac{\partial^2\gamma}{\pi}\exp[-\frac{2}{T}|\zeta-\gamma e^{-\kappa t}|^2]W(\gamma,\gamma^*,0)
\end{equation}
with $T = 1- \exp(-2\kappa t)$, $W(\gamma,\gamma^*,0)$ is the Wigner function at initial time $t=0$ which is given in \eqref{ewf}. The time evolution of the Wigner function \eqref{dwlc} models dissipation due to interaction with a vacuum reservoir as well as inefficient detectors with efficiency $\eta=1-T$ . The detailed analytic expression for \eqref{dwlc} is obtained in Appendix~\ref{adcwlc}.
\begin{figure*}[hbt]
\includegraphics[scale=.9]{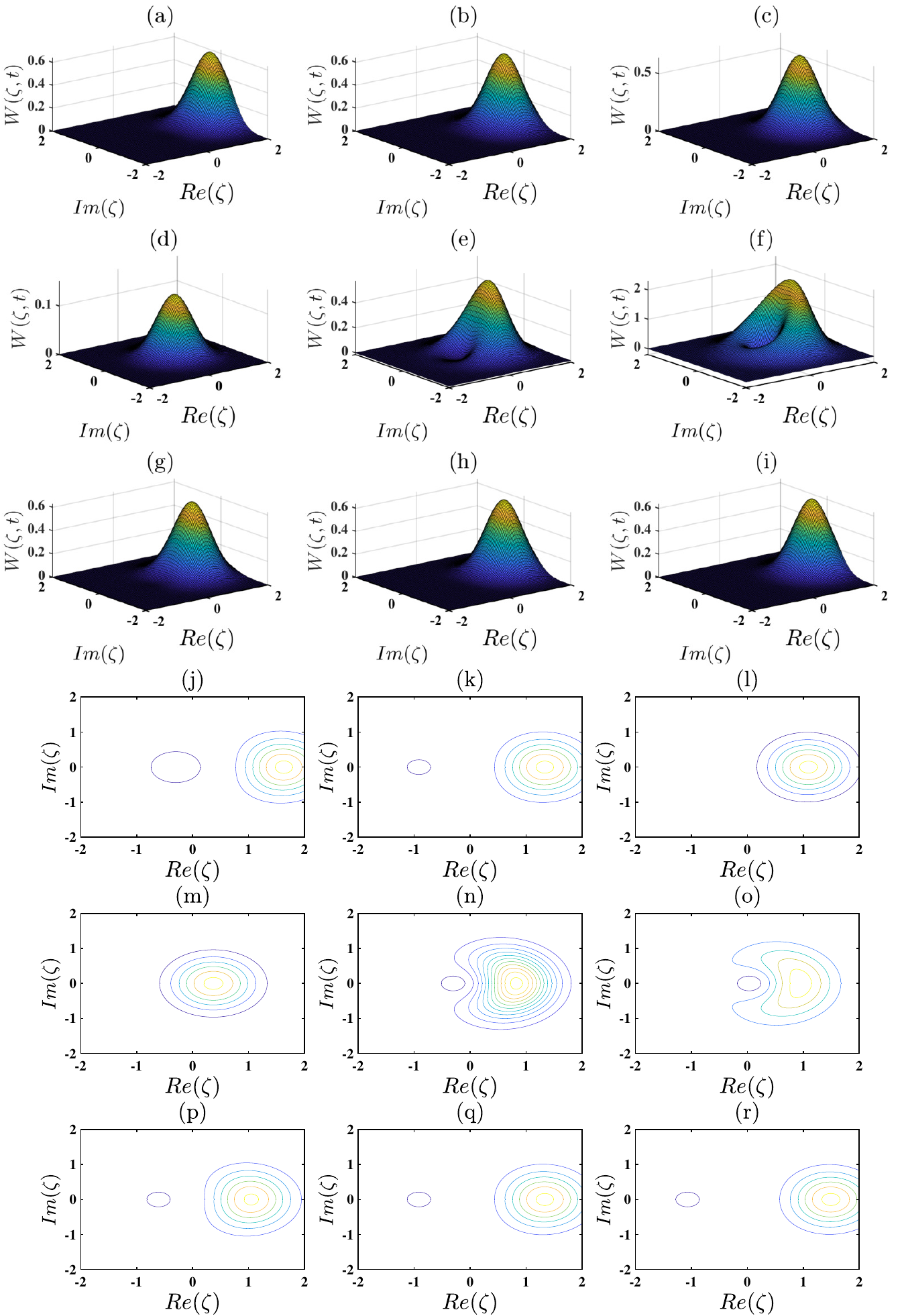}
\caption{A comparison of Wigner function w.r.t. real and imaginary $\zeta$ for $\alpha = 1.5,\,t=1/\sqrt{2}$ and (a) $\kappa t=0.1$, (b) $\kappa t=0.3$, (c) $\kappa t=0.5$; for $\kappa t=0.3,\,\alpha=0.5$ and (d) $t= 1$, (e) $t = 1/\sqrt{2}$, (f) $t = 0$; for $\kappa t=0.3,\,t=1/\sqrt{2}$ and (g) $\alpha= 1$, (h) $ \alpha= 1.5$, (i) $\alpha=1.75$, (j) to (r) are the corresponding contour plots.}
\label{figwlc}
\end{figure*}
It is clear from Fig.~\ref{figwlc} that with an increase in the rescaled time $\kappa t$, non-Gaussianity decreases for fixed values of $\alpha$ and $t$ and nonclassicality nature is not revealed at all. Similarly, with an increase (decrease) in $r$ $(t)$, non-Gaussianity nature increases, and nonclassicality is not detected. Also, by increasing $\alpha$ with fixed $t$ and $\kappa t$ there are merely any changes in non-Gaussianity.
\section{Conclusion}
\label{conclsn}
From the above discussion, we can conclude that the Wigner function detects the non-Gaussianity as well as the nonclassicality of CSQS. But if the coherent superposed quantum state is evolved via a photon loss channel, the Wigner function shows the non-Gaussianity but fails to detect the nonclassicality. Linear entropy decreases with an increase in coherent state parameter $\alpha$ and increases with increasing $r$. $N(\rho)$ also has a similar behavior as linear entropy. Wigner logarithmic negativity decreases with $\alpha$ up to a fixed value of $\alpha$ and then increases. Also, it is negative in most of the regions and increases with an increase in the value of $r$. The relative entropy measure first decreases with $\alpha$ up to a fixed value and then increases with $r$. Thus the considered coherent superposed quantum state is nonclassical as well as non-Gaussian. 

The results reported here seem to be experimentally realizable, as photon-added coherent states are prepared in laboratory by Bellini et. al. \cite{png64} and non-Gaussianity present in this state is also measured experimentally \cite{dp2}. Keeping this in mind, we end the article with an expectation that the current result will be of further use for performing quantum optical and computing
tasks.

\begin{center}
\textbf{ACKNOWLEDGEMENT}
\end{center}

Deepak’s work is supported by the Council of Scientific and Industrial Research
(CSIR), Govt. of India (Award no. 09/1256(0006)/2019-EMR-1).

\begin{widetext}
\begin{appendix}
\section{Linear Entropy}
\label{aep}
The density matrix using \eqref{phiep} is given by
\begin{eqnarray*}
\rho_{AB} & = &\ket{\phi}_{AB} {}_{AB}\bra{\phi}\\
& = & \sum_{p=0}^\infty \left(c_{1p}\sum_{j=0}^p\sqrt{p\choose j}\ket{j,p-j}+c_{2p}\sum_{j=0}^{p+1}\sqrt{p+1\choose j}\ket{j,p+1-j}\right)\times\\
& & \sum_{q=0}^\infty \left(c_{1q}^*\sum_{k=0}^q\sqrt{q\choose k}\bra{k,q-k}+c_{2q}^*\sum_{k=0}^{q+1}\sqrt{q+1\choose k}\bra{k,q+1-k}\right)\\
\Rightarrow \rho_B & = &
\sum_{p,q=0}^\infty \left(c_{1p}c_{1q}^*\sum_{j=0}^{p}\sqrt{{p\choose j}{q\choose j}}\ket{p-j}\bra{q-j} + c_{1p}c_{2q}^*\sum_{j=0}^{p}\sqrt{{p\choose j}{q+1\choose j}}\ket{p-j}\bra{q+1-j}
\right.\\
& & + \left.c_{2p}c_{1q}^*\sum_{j=0}^{p+1}\sqrt{{p+1\choose j}{q\choose j}}\ket{p+1-j}\bra{q-j} + c_{2p}c_{2q}^*\sum_{j=0}^{p+1}\sqrt{{p+1\choose j}{q+1\choose j}}\ket{p+1-j}\bra{q+1-j} \right) \\
\Rightarrow \text{Tr}(\rho_B^2) & = &
N^4\left[t^4|\alpha|^4 + t^3s\left\{2\alpha^2|\alpha|^2 + 2\alpha^{*2}|\alpha|^2 \right\} + t^2s^2 \left\{2|\alpha|^2(1+2|\alpha|^2) + \alpha^4 + \alpha^{*4} \right\} \right. \\ & & + \left.
ts^3\left\{2\alpha^2(1+|\alpha|^2) + 2\alpha^{*2}(1+|\alpha|^2)\right\} + s^4(1+5|\alpha|^2+2|\alpha|^4)/2 \right]
\end{eqnarray*}

\section{Wigner Logarithmic Negativity}
\label{awln}

\begin{align*}
W & = log_2\left(\int{d^2\gamma}|W(\gamma,\gamma^*)|\right) \\ & =
log_2\left(\int{d^2\alpha}|W_0(\alpha)N^2[|t\alpha_0+r(2\alpha^*-\alpha_0^*)|^2-|r|^2]|\right) \\
& =
log_2\left(\frac{2N^2}{\pi}\int{d^2\alpha}\exp(-2|\alpha-k|^2)\left||tk+r(2\alpha^*-k)|^2-|r|^2\right|\right)~~~(\text{assuming $\alpha_0=k$, a real number for simplicity}) \\ & =
log_2\left(\frac{2N^2}{\pi}\int{d^2\alpha}\exp(-2|\alpha-k|^2)\left|(t-r)^2k^2+4r^2|\alpha|^2+2r(t-r)k(\alpha+\alpha^*)-r^2\right|\right)~(\text{putting $\alpha=x+iy$}) \\ & =
log_2\left(\frac{2N^2}{\pi}\int{dxdy}\exp(-2(x-k)^2-2y^2)\left|(t-r)^2k^2+4r^2(x^2+y^2)+4r(t-r)kx-r^2\right|\right)
\end{align*}
Let $x_1$ and $x_2$ be the two solutions of $4r^2x^2+4r(t-r)kx+(t-r)^2k^2+4r^2y^2-r^2$ with $x_1\leq x_2=0$ and hence we get
$$x_1=(-(t-r)k-r\sqrt{r^2-4y^2})/2r$$
$$x_2=(-(t-r)k+r\sqrt{r^2-4y^2})/2r$$
Thus the modulus inside the integration is positive on the intervals $(-\infty,x_1)$ and $(x_2,\infty)$ and negative in $(x_1,x_2)$.
Using $\int{dx}\exp(-2x^2)=\sqrt{\frac{\pi}{2}}$, $\int{dx}\exp(-2x^2)x=0$, $\int{dx}\exp(-2x^2)x^2=\frac{3}{4}\sqrt{\frac{\pi}{2}}$, we can calculate as
\begin{align*}
W & = log_2\left[N^2\left((t+r)^2k^2-5r^2 \right)
 \right]
\end{align*}
\section{Wigner function evolving under the photon loss channel}
\label{adcwlc}
\begin{align*}
W(\zeta,t)& = \frac{2}{T}\int\frac{d^2\alpha}{\pi}\exp[\frac{-2}{T}|\zeta-\alpha e^{-\kappa t}|^2]\frac{2N^2}{\pi}\left[|\alpha_0|^2+r^2\left(-1+4|\alpha|^2-2\alpha\alpha_0^*-2\alpha^*\alpha_0\right) \right.\\&+\left. rt\left(2\alpha\alpha_0+2\alpha^*\alpha_0^*-\alpha_0^{*2}-\alpha_0^2\right)\right] \exp(-2|\alpha-\alpha_0|^2) \\
& =
\frac{4N^2}{T\pi}\int\frac{d^2\alpha}{\pi}\exp[\frac{-2}{T}\left(|\zeta|^2+T|\alpha_0|^2+|\alpha|^2\{T+e^{-2\kappa t}\}-\alpha(\zeta^*e^{-\kappa t}+T\alpha_0^*)-\alpha^*(\zeta e^{-\kappa t}+T\alpha_0)\right)]\\ &\times\left[|\alpha_0|^2+r^2\left(-1+4|\alpha|^2-2\alpha\alpha_0^*-2\alpha^*\alpha_0\right) + rt\left(2\alpha\alpha_0+2\alpha^*\alpha_0^*-\alpha_0^{*2}-\alpha_0^2\right)\right]\\
& (\text{using $T+e^{-2\kappa t}=1$ and letting $\zeta e^{-\kappa t}+T\alpha_0=\eta$}) \\
& =
\frac{2N^2}{\pi}\exp[\frac{2}{T}\left(|\eta|^2-|\zeta|^2-T|\alpha_0|^2\right)]\left[t^2|\alpha_0|^2+r^2\left(-1+2T+|2\eta-\alpha_0|^2\right) + rt\left(2\text{Re}[(2\eta-\alpha_0)\alpha_0]\right)\right]
\end{align*}
\end{appendix}
\end{widetext}

\end{document}